\documentstyle[prl,aps,epsf,multicol]{revtex}
\begin{document}
\draft
\title{Phase diffusion and charging effects in Josephson junctions}
\author{Hermann Grabert}
\address{Fakult\"at f\"ur Physik, Albert-Ludwigs-Universit{\"a}t,
Hermann-Herder-Stra{\ss}e~3, D-79104 Freiburg, Germany}
\author{Gert-Ludwig Ingold}
\address{Institut f\"ur Physik, Universit\"at Augsburg,
Memminger Stra{\ss}e~6, D-86135 Augsburg, Germany}
\author{Benjamin Paul}
\address{Fachbereich Physik, Martin-Luther-Universit\"at, 
Selkestra{\ss}e~9E,  D-06099 Halle, Germany}
%\date{January 16, 1998}
\maketitle
\widetext
\begin{abstract}
The supercurrent of a Josephson junction is reduced by phase diffusion. 
For ultrasmall capacitance junctions the current may be further decreased 
by Coulomb blockade effects. 
We calculate the Cooper pair current by means of 
time-dependent perturbation theory to all orders in the Josephson coupling 
energy and obtain the current-voltage characteristic in closed
form in a range of parameters of experimental interest. 
The results comprehend phase diffusion of the coherent Josephson
current in the classical regime as well as the supercurrent peak
due to incoherent Cooper pair tunneling in the strong Coulomb 
blockade regime.
\end{abstract}


\raggedcolumns
\begin{multicols}{2}
\narrowtext
New lithography and low-temperature techniques have allowed the
fabrication and measurement of small Josephson junctions affected
by the capacitive charging energy of single Cooper 
pairs\cite{schoe90,sct91,tinkh96}.
Much of the work so far has concentrated on the region of
strong Coulomb blockade where the tunneling of Cooper pairs described
by the Josephson energy can be treated perturbatively.
While Coulomb blockade of Cooper pair tunneling is fairly well understood,
the relation between the effects observed at low temperatures
and the familiar ``classical'' dynamics of Josephson
junctions\cite{baron82} remains to be exemplified. In this
article we demonstrate that the current peak caused by
incoherent Cooper pair tunneling in the regime of strong
Coulomb blockade gradually evolves into the classical supercurrent
when parameters are changed accordingly. 

\begin{figure}
\begin{center}
\leavevmode
\epsfxsize=0.35\textwidth
\epsfbox{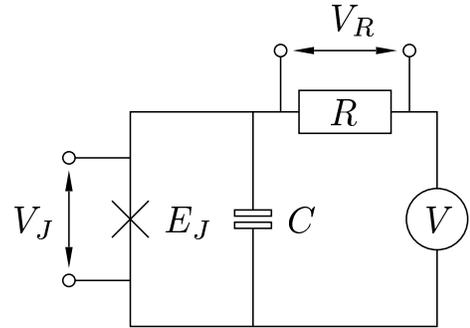}
\end{center}
\caption{A Josephson junction characterized by the Josephson energy
$E_J$ and capacitance $C$ is coupled to a voltage source $V$ via a resistor
$R$. The voltage drops across the junction and the resistor are $V_J$ and
$V_R$, respectively.}
\label{fig:fig1}
\end{figure}

We consider a Josephson junction with capacitance $C$ and critical 
current $I_c=(2e/\hbar)E_J$, where $E_J$ is the Josephson energy. The 
capacitance gives rise to a charging energy $E_c=2e^2/C$ for Cooper pairs. 
The junction is coupled to an ideal voltage source through a resistor 
of resistance $R$ as shown in Fig.~\ref{fig:fig1}. 
This system can be modeled by the Hamiltonian
\begin{equation}
H=H_J+H_{\rm env}
\label{eq:hamil}
\end{equation}
where the first term
\begin{equation}
H_J=-E_J\cos(\varphi).
\label{eq:hj}
\end{equation}
is related to the tunneling of Cooper pairs through the Josephson junction.
In analogy to the familiar relation between the voltage across the Josephson 
junction and the corresponding phase difference
\begin{equation}
V_J=\frac{\hbar}{2e}\dot\varphi,
\label{eq:vphi}
\end{equation}
we may introduce a phase 
\begin{equation}
\varphi_R=\frac{2e}{\hbar}Vt-\varphi
\label{eq:phir}
\end{equation}
related to the voltage $V_R=V-V_J$ across the resistor where $V$ is the external
voltage. In the following, we are interested in a voltage regime far below the 
gap voltage so that tunneling of quasiparticles may be neglected. 

The second part of the Hamiltonian (\ref{eq:hamil}) describing the coupling of 
the junction to its electrodynamic environment 
reads\cite{ingol91}
\begin{equation}
H_{\rm env}=\frac{Q^2}{2C}+\sum_{n=1}^{\infty}\left[\frac{q_n^2}{2C_n}+
\left(\frac{\hbar}{2e}\right)^2\frac{1}{2L_n}(\varphi_R-\varphi_n)^2\right].
\label{eq:hr}
\end{equation}
The first term represents the charging energy where the 
charge $Q$ obeys the commutation relation $[\varphi,Q]=2ie$.
By virtue of Eq.\ (\ref{eq:phir}), the second part describes a bilinear\break
\noindent
coupling between the phase difference across the Josephson junction and 
phases of a set of $LC$-circuits which model 
the electromagnetic environment of the junction. The phases $\varphi_n$ and 
charges $q_n$ are conjugate variables according to $[\varphi_n,q_n]=2ie$. It 
can be shown that by suitably choosing the parameters $C_n$ and $L_n$, the 
second part of $H_{\rm env}$ effectively describes the resistor $R$.

The thermal equilibrium Cooper pair current at temperature $T=1/k_B\beta$
may be written as
\begin{equation}
I_S = I_c\left\langle
{\cal U}^+(\infty,t_0)\sin[\varphi(t_0)]{\cal U}(\infty,t_0)
\right\rangle_{\beta}
\label{eq:is0}
\end{equation}
where the time dependence of $\varphi(t)$ arises from the Hamiltonian
(\ref{eq:hr}).
\begin{equation}
{\cal U}(t,t_0)={\cal T}\exp\left(\frac{i}{\hbar}\int_{t_0}^{t}dt' E_J
\cos[\varphi(t')]\right)  
\label{eq:u}
\end{equation}
with the time ordering operator ${\cal T}$ is the time evolution operator
in the interaction representation. Taking the limit
$t\to\infty$, the precise value of the initial time $t_0$ becomes irrelevant
and will be set to zero in the following. Expanding the current (\ref{eq:is0})
in powers of the Josephson energy $E_J$ one obtains
\begin{eqnarray}
I_S&=&i\frac{I_c}{2}\sum_{M=1}^{\infty}\left(
\frac{i}{2\hbar}E_J\right)^{2M-1}\sum_{\{\zeta,\eta\}}\left(\prod_{k=1}^{2M-1}
\eta_k\right) \zeta_0\nonumber\\
&&\times\int_0^{\infty}dt_1\dots\int_0^{t_{2M-2}}dt_{2M-1}\nonumber\\
&&\times\exp\left(i\frac{2e}{\hbar}V\sum_{k=0}^{2M-1}\zeta_kt_k\right)
\nonumber\\
&&\times\exp\left(-\sum_{k=1}^{2M-1}\sum_{l=0}^{k-1}
\zeta_k\zeta_lJ[\eta_k(t_k-t_l)]\right).
\label{eq:is1}
\end{eqnarray}
For given order $M$, the $\zeta_k, k=0,\dots, 2M-1 $, which arise from a 
decomposition of the trigonometric functions in Eqs.\ (\ref{eq:is0}) and 
(\ref{eq:u}) into exponentials, may take the values $\pm1$ and are subject to 
the constraint $\sum_{k=0}^{2M-1}\zeta_k=0$. The factors $\eta_k$ also take 
the values $\pm1$ and account for the fact that Eq.\ (\ref{eq:is0}) contains 
two time-ordered operators. Finally, the phase correlation 
function\cite{ingol91} 
\begin{equation}
J(t)=2\int_{-\infty}^{+\infty}\frac{d\omega}{\omega}\frac{{\rm Re}Z_t(\omega)}
{R_Q}\frac{e^{-i\omega t}-1}{1-e^{-\beta\hbar\omega}}
\label{eq:joft}
\end{equation}
contains the relevant information on the total environmental impedance. 

In our case, the impedance as seen from the tunnel junction contains a 
capacitance $C$ in parallel with a resistance $R$ leading to
\begin{equation}
\frac{{\rm Re}Z_t(\omega)}{R_Q}=\frac{\rho}{1+(\omega/\omega_R)^2}.
\label{eq:rezt}
\end{equation}
The dimensionless impedance at zero frequency is $\rho=R/R_Q$ with the 
resistance quantum $R_Q=h/4e^2$, while the frequency scale is set by 
$\omega_R=1/RC$.

For the impedance (\ref{eq:rezt}), evaluation of the integral (\ref{eq:joft}) 
gives
\begin{eqnarray}
J(t)=-2\rho\biggl[&&\frac{\pi}{\hbar\beta}\vert t\vert
+S +\frac{\pi}{2} e^{-\omega_R \vert t\vert}
\cot\left(\frac{\beta\hbar\omega_R}{2}\right) \nonumber\\
&&-\sum_{n=1}^{\infty}\frac{e^{-\nu_n\vert t\vert}}{n[1-
(\nu_n/\omega_R)^2]}\nonumber\\
&&+i\frac{\pi}{2}(1-e^{-\omega_R\vert t\vert}){\rm sgn}(t)\biggr]
\label{eq:joftoe}
\end{eqnarray}
with
\begin{equation}
S=\gamma+\frac{\pi^2\rho}{\beta E_c}+\psi\left(\frac{\beta E_c}{2\pi^2\rho}
\right).
\label{eq:upsilon}
\end{equation}
Here, $\gamma=0.5772\dots$ is Euler's constant and the 
$\nu_n=2\pi n/\hbar\beta$ are the Matsubara frequencies.
Further, $\psi(x)$ is the logarithmic derivative of the gamma function and 
${\rm sgn}(x)$ denotes the signum function. 

Now, typical lead resistances are of the order of 100$\Omega$\cite{devor91}
so that $\rho\ll 1$. Then, for small Josephson junctions the large
lead conductance overdamps the junction and $\omega_R$ is large compared to 
the Josephson frequency
\begin{equation}
\omega_J=\frac{2e}{\hbar}RI_c.
\end{equation}
Hence, we may disregard terms proportional to $\exp(-\omega_Rt)$,
and Eq.\ (\ref{eq:joftoe}) reduces to
\begin{eqnarray}
J(t)=-2\rho\biggl[&&\frac{\pi}{\hbar\beta}\vert t\vert+
S -\sum_{n=1}^{\infty}\frac{e^{-\nu_n\vert
t\vert}}{n[1-(\nu_n/\omega_R)^2]}\nonumber\\
&&+i\frac{\pi}{2}{\rm sgn}(t)\biggr].
\label{eq:joftod}
\end{eqnarray}
Furthermore, since $\rho\ll1$, the ratio $\nu_1/\omega_J=1/\rho\beta E_J$ may 
be large even for low temperatures, and the result (\ref{eq:joftod}) can be 
simplified further to read
\begin{equation}
J(t)=-2\rho\left[\frac{\pi}{\hbar\beta}\vert t\vert+S
+i\frac{\pi}{2}{\rm sgn}(t)\right].
\label{eq:joftlt}
\end{equation}
Inserting this correlation function into (\ref{eq:is1}), the time integrals may
be performed and one obtains the expansion
\begin{equation}
I_S = I_c{\rm Re}\left(\sum_{M=1}^{\infty}
\sum_{\{x_k\}}\prod_{k=1}^{2M-1}a(x_k)\right),
\label{eq:is2}
\end{equation}
where the set of positive integers $\{x_k\}$ describes a sequence with
$\vert x_{k+1}-x_k\vert=1$ starting and
ending at $x_1=x_{2M-1}=1$. Each value of $x_k$ is associated with a factor
\begin{equation}
a(x_k)=\frac{\sin(\pi\rho x_k)}{2\pi\rho x_k}
\frac{\exp[2\rho S x_k(x_{k+1}-x_k)]}
{\nu+ix_k/\beta E_J}
\label{eq:aofx}
\end{equation}
which contains the dependence of the Cooper pair current on the applied voltage
$V$ via the dimensionless variable $\nu=V/RI_c$. An expansion of the form
(\ref{eq:is2}) appears in the context of the one-dimensional Coulomb gas 
\cite{lenar61} and was also found in a previous treatment of the dynamics
of small current-biased Josephson junctions \cite{zwerg87} 
for a different kind of approximation. Following \cite{lenar61}, we may rewrite
(\ref{eq:is2}) in terms of a continued fraction 
\begin{equation}
I_S = I_c{\rm Re}\left[\frac{\sin(\pi\rho)}
{2\pi\rho}\frac{\exp(-2\rho S)}{\nu+i/\beta E_J}
\displaystyle\frac{1}{1+\displaystyle\frac{b_1}{1+
\displaystyle\frac{b_2}{1+\dots}}}\right] 
\label{eq:is3}
\end{equation}
with coefficients
\begin{equation}
b_n=\left(\frac{\beta E_J}{2\pi\rho}\right)^2\frac{\sin(\pi\rho n)
\sin(\pi\rho(n+1)) \exp(-2\rho S)}
{n(n+1)(n-i\nu\beta E_J)(n+1-i\nu\beta E_J)}.
\label{eq:cfcoeff}
\end{equation}

It is instructive to first investigate the limit $\rho\to 0$ where the 
coefficients simplify to read
\begin{equation}
b_n=\left(\frac{\beta E_J}{2}\right)^2\frac{1}
{(n-i\nu\beta E_J)(n+1-i\nu\beta E_J)}
\label{eq:cfcoeffcl}
\end{equation}
and the continued fraction (\ref{eq:is3}) may 
be evaluated by means of a matrix recursion \cite{abram72}. The 
recursion relations can be solved in terms of modified Bessel functions of 
complex order. For finite $\rho\ll 1$ the corrections are not necessarily 
small since in the current-voltage characteristic (\ref{eq:is3}) 
$\rho$ is multiplied by a factor $S$ defined in Eq.\ (\ref{eq:upsilon}) which 
may contain large terms of order $\log(\beta E_c/\rho)$. 
On the other hand, a numerical evaluation of the continued 
fraction converges rapidly and only coefficients $b_n$ with moderately large 
$n$ are relevant. Thus, for $\rho\ll 1$, the sine functions in Eq.\ 
(\ref{eq:cfcoeff}) may be linearized and it is readily seen that the dominant 
corrections are described in terms of an effective Josephson energy
\begin{equation}
E_J^*=E_J\exp(-\rho S).
\label{eq:ejeff}
\end{equation}
The continued fraction may then be evaluated by the method described above
leading to the current-voltage characteristic
\begin{equation}
I_S=\frac{2e}{\hbar}E_J^*\,{\rm Im}\left(\frac{I_{1-i\beta eV/\pi\rho}(\beta
E_J^*)}{I_{-i\beta eV/\pi\rho}(\beta E_J^*)}\right).
\label{eq:genivanzilb}
\end{equation}
This constitutes the central result of this paper. The dependence of the
effective Josephson energy on temperature is depicted in Fig.~2 for 
various values of $\rho$.

To investigate the classical limit of the current-voltage characteristic 
(\ref{eq:genivanzilb}) one must know that the limit $\hbar\to 0$\break
\begin{figure}
\begin{center}
\leavevmode
\epsfxsize=0.45\textwidth
\epsfbox{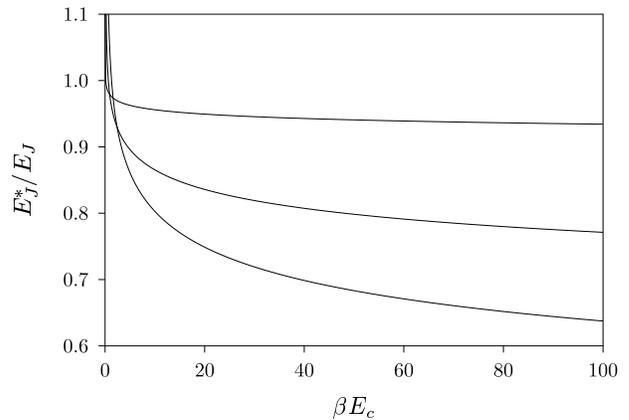}
\end{center}
\caption{The effective Josephson energy $E_J^*$ is shown
as a function of the inverse temperature for $\rho = 0.01, 0.05,$ and 0.1 from
the upper to the lower curve.}
\label{fig:fig2}
\end{figure}
\noindent
has to be performed such that the flux quantum $h/2e$ remains constant. 
This means that $\rho$ is of order $\hbar$ and $E_c$ of order $\hbar^2$. 
Then, in the overdamped limit $\omega_R\gg\omega_J$ the effective Josephson 
energy coincides with the bare Josephson energy $E_J$. This can also be seen 
in Fig.~\ref{fig:fig2} where for small $\rho$ the range in which $E_J^*$ 
practically equals $E_J$ becomes very large. In this case, the current-voltage 
characteristic (\ref{eq:genivanzilb}) can be shown to reduce to the standard 
Ivanchenko-Zil'berman result\cite{ivanc69,ingol94b} for classical overdamped 
Josephson junctions.

While in the classical limit $\beta E_c/\rho$ vanishes,
in the limit of strong Coulomb blockade we have $\beta E_c/\rho\gg 1$.
The psi function in Eq.\ (\ref{eq:upsilon}) may then be approximated 
by a logarithm to obtain 
\begin{equation}
S=\gamma+\ln(\beta E_c/2\pi^2\rho) .
\label{eq:ssimpl}
\end{equation} 
Current-voltage characteristics for overdamped ultrasmall Josephson
junctions have so far only been calculated to lowest order in the Josephson
energy. The supercurrent arising from incoherent
Cooper pair tunneling then reads\cite{ingol94b,grabe93,ingol94a}
\begin{equation}
I_S=f\,\frac{\vert \Gamma(\rho-i\beta eV/\pi)\vert^2}{\Gamma(2\rho)}\sinh(\beta
eV)
\label{eq:scpeak}
\end{equation}
where
\begin{equation}
f=\frac{\pi e}{\hbar}\frac{E_J^2}{E_c}\rho^{2\rho}\left(\frac{\beta E_c}
{2\pi^2}\right)^{1-2\rho}\exp(-2\rho\gamma).
\label{eq:prefscp}
\end{equation}

\begin{figure}
\begin{center}
\leavevmode
\epsfxsize=0.45\textwidth
\epsfbox{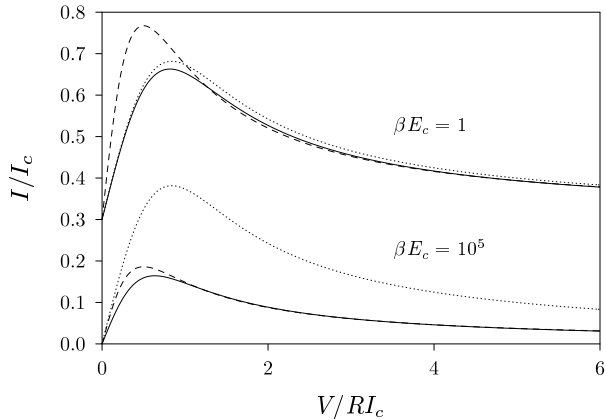}
\end{center}
\caption{The current-voltage characteristic of a Josephson junction
with Josephson energy $\beta E_J=2$ is shown for charging energies
$\beta E_c=1$ and $10^5$ and external resistance $\rho=0.04$. The full line
corresponds to our result (\protect\ref{eq:genivanzilb}) while the dotted line
gives the standard Ivanchenko-Zil'berman result and
the dashed line depicts the prediction (\protect{\ref{eq:scpeak}}) for strong 
Coulomb blockade. The two current-voltage characteristics are
vertically shifted with respect to each other by $I/I_c=0.3$ for sake of
clarity.}
\label{fig:fig3}
\end{figure}

In Fig.~\ref{fig:fig3}, we compare our result (\ref{eq:genivanzilb}) (full line)
with the standard Ivanchenko-Zil'berman result (dotted line) and the Coulomb
blockade result (\ref{eq:scpeak}) (dashed line) for $\beta E_J=1$, $\rho=0.04$,
and two different values for $\beta E_c$. For large $\beta E_c$, the 
current-voltage characteristic (\ref{eq:genivanzilb}) is
in very good agreement with the Coulomb blockade result (\ref{eq:scpeak}). With 
decreasing $\beta E_c$, the results (\ref{eq:genivanzilb}) and\break
\noindent
(\ref{eq:scpeak}) evolve differently and a crossover to the 
Ivanchenko-Zil'berman result can be found. 

To make the connection between our result (\ref{eq:genivanzilb}) and the
Coulomb blockade result (\ref{eq:scpeak}) more explicit, we consider the 
current-voltage characteristic (\ref{eq:genivanzilb}) to order $E_J^2$ using 
the approximation (\ref{eq:ssimpl}). We then obtain for the Cooper pair 
current
\begin{equation}
I_S=f \frac{2\pi^2\rho\beta eV}{(\beta eV)^2+\pi^2\rho^2}.
\label{eq:is4}
\end{equation}
While the approximation of the phase correlation function $J(t)$ leading 
from Eq.\ (\ref{eq:joftod}) to Eq.\ (\ref{eq:joftlt}) is only justified for 
temperatures with $\rho\beta E_J\ll 1$, for $E_J\ll E_c$ the theory should 
extend to rather low temperatures. In fact, the zero 
bias differential resistance as well as the voltage for which the current takes
its maximum are obtained precisely for small $\rho$ including terms of order 
$\rho^2$.

The result (\ref{eq:is4}) may be improved by going beyond the approximation
(\ref{eq:joftlt}) for the phase correlation function $J(t)$. Keeping the sum
over the Matsubara frequencies in Eq.\ (\ref{eq:joftod}), 
a closed form for $J(t)$
can be found in the strong Coulomb blockade regime $\beta E_c/\rho\gg 1$. The 
current-voltage characteristic (\ref{eq:is1}) may then be evaluated to order
$E_J^2$ and reproduces exactly the expression (\ref{eq:scpeak}).

In summary, we have derived a formally exact expansion of the
current-voltage characteristic of a voltage-biased
Josephson junction as a power series in the Josephson coupling
energy $E_J$. Noting that ultrasmall junctions in  a standard electromagnetic
environment are overdamped, i.e., $\omega_J/\omega_R=(2e/\hbar)R^2I_cC\ll1$,
and with the assumption $\rho\beta E_J=(eRI_c/\pi k_BT)\ll1$,
the general result was written as continued fraction, which 
for $\rho\ll1$ could be summed in closed form. The inferred main
result (\ref{eq:genivanzilb}) was shown to describe the changeover from
the classical Josephson effect in the presence of phase diffusion
to the quantum regime of strong Coulomb blockade where Cooper
pairs tunnel incoherently. The theory covers the experimentally 
relevant range for small capacitance Josephson junctions. Although
detailed experimental studies of the region between the
above-mentioned limits are absent, recent work\cite{vion96}
indicates quantum effects in qualitative accord with the predictions made.

We would like to thank the authors of Ref.~\cite{vion96} for stimulating
discussions on experimental aspects of the problem and acknowledge helpful
conversations with G.\ G\"oppert. Financial support was provided by the
Deutsche Forschungsgemeinschaft.

\end{multicols}
\end{document}